\documentclass[conference]{IEEEtran}
\IEEEoverridecommandlockouts
\usepackage[T1]{fontenc}
\usepackage{graphicx}
\usepackage{url}
\usepackage{amssymb}
\usepackage{amsthm}
\usepackage{amsmath}
\DeclareMathOperator*{\argmax}{arg\,max}

\newtheorem{dfn}{Definition}
\newtheorem{question}{Question}
\newtheorem{theorem}{Theorem}
\newtheorem{example}{Example}

\newcommand{\probP}{\text{I\kern-0.15em P}}
\usepackage{mathtools}

\DeclarePairedDelimiter\floor{\lfloor}{\rfloor}
\usepackage{cite}
\newenvironment{manualtheorem}[1]{%
  \renewcommand\thetheorem{#1}%
  \begin{theorem}%
  \addtocounter{theorem}{-1}%
}{%
  \end{theorem}%
}
\def\BibTeX{{\rm B\kern-.05em{\sc i\kern-.025em b}\kern-.08em
    T\kern-.1667em\lower.7ex\hbox{E}\kern-.125emX}}

\newcommand{\myscale}[2][0.7]{% Default scale is 0.8
    \scalebox{#1}{$#2$}
}

\begin{document}

\title{Opportunity-Cost-Driven Reward Mechanisms for Crowd-Sourced Computing Platforms}

\author{
    \IEEEauthorblockN{
        Shuhao Zheng\textsuperscript{*}\IEEEauthorrefmark{2},
        Ziyue Xin\textsuperscript{*}\IEEEauthorrefmark{2},
        Zonglun Li\IEEEauthorrefmark{2},
        Xue Liu\IEEEauthorrefmark{2}\IEEEauthorrefmark{3}
    }
    \IEEEauthorblockA{
        \IEEEauthorrefmark{2}McGill University\\
        \IEEEauthorrefmark{3}Mohamed Bin Zayed University of Artificial Intelligence\\
        \{shuhao.zheng, ziyue.xin, zonglun.li\}@mail.mcgill.ca\\
        xue.liu@cs.mcgill.ca
    }
    % \IEEEauthorblockA{

    % }
    % \IEEEauthorblockA{
    %     \IEEEauthorrefmark{3}Affiliation 3\\
    %     Email: third.author@example.com
    % }
    % \IEEEauthorblockA{
    %     \IEEEauthorrefmark{4}Affiliation 4\\
    %     Email: fourth.author@example.com
    % }
    \thanks{\textsuperscript{*}These authors contributed equally to this work.}
}

\maketitle

\begin{abstract}
This paper introduces a game-theoretic model tailored for reward distribution on crowd-sourced computing platforms. It explores a repeated game framework where miners, as computation providers, decide their computation power contribution in each round, guided by the platform's designed reward distribution mechanism. The reward for each miner in every round is based on the platform's randomized task payments and the miners' computation transcripts. Specifically, it defines Opportunity-Cost-Driven Incentive Compatibility (OCD-IC) and Dynamic OCD-IC (DOCD-IC) for scenarios where strategic miners might allocate some computation power to more profitable activities, such as Bitcoin mining. The platform must also achieve Budget Balance (BB), aiming for a non-negative total income over the long term. This paper demonstrates that traditional Pay-Per-Share (PPS) reward schemes require assumptions about task demand and miners' opportunity costs to ensure OCD-IC and BB, yet they fail to satisfy DOCD-IC. The paper then introduces Pay-Per-Share with Subsidy (PPSS), a new reward mechanism that allows the platform to provide subsidies to miners, thus eliminating the need for assumptions on opportunity cost to achieve OCD-IC, DOCD-IC, and long-term BB.
\end{abstract}

\section{Introduction}
The rapid expansion of modern technology necessitates a significant amount of computation, spanning fields such as mathematics, biology, astrophysics, and artificial intelligence. Centralized computing solutions, including cloud services and supercomputers, face limitations in availability and resources, often coming with a high rental cost. Meanwhile, the widespread availability of personal computers and private data centers has led to the rise of crowd-sourced computing as an efficient and cost-effective method to meet the extensive computation needs by pooling resources globally. Crowd-sourced computation initiatives like Einstein@Home\footnote{\url{https://einsteinathome.org/}} and PrimeGrid\footnote{\url{https://www.primegrid.com/}} have successfully harnessed a vast amount of volunteered computation power, thanks to the development of significant shared-computing frameworks such as BOINC \cite{andersonBOINCPlatformVolunteer2020}. Nevertheless, maintaining a steady flow of volunteer contributions is challenging without adequate incentives, given the costs associated with electricity and maintenance.

Recent years have seen the rise of incentive-driven crowd-sourced platforms, notably Bitcoin mining pools \cite{IncentiveCompatibilityBitcoin} and crowd-sourced storage services like Filecoin\footnote{\url{https://filecoin.io/}}. Schrijvers et al. \cite{IncentiveCompatibilityBitcoin} conceptualized the allocation problem within a Bitcoin mining pool as a reward mechanism, introducing the pay-per-last-N-shares function that achieves both incentive compatibility and budget balance. This design presumes the existence of a singular Bitcoin mining pool, simplifying miners' decision-making to whether they should delay the reporting of successfully mined blocks. However, crowd-sourced computing platforms in the real world encounter a prevalent challenge: strategic miners might allocate a portion of their computation power to alternative profitable ventures, such as Bitcoin mining, if they perceive higher gains, thereby confronting the concept of opportunity cost. The core issue then revolves around modeling miners' strategies under these conditions and crafting a reward mechanism, potentially with subsidies, to motivate miners to dedicate their entire computation power to the crowd-sourced platform while ensuring the platform's budget balance for sustainability.

This paper frames the reward distribution on a crowd-sourced computing platform as a repeated decision game. In each round, miners independently decide what percentage of their computation power to allocate to the platform. The objective is to devise a reward mechanism that distributes rewards to miners, taking into account the task payments, the computation transcripts, and the miners' opportunity costs. Furthermore, this paper introduces the concepts of Opportunity-Cost-Driven Incentive Compatibility (OCD-IC) and Dynamic Opportunity-Cost-Driven Incentive Compatibility (DOCD-IC), aiming for miners to commit the entirety of their computation power to the platform as they report. At the same time, it's crucial for the platform to achieve Budget Balance for sustainability. Given the variability of task payments in each round, the platform might have to offer subsidies to miners alongside payments to meet (D)OCD-IC requirements. Therefore, Budget Balance (BB) in this context is envisioned as a long-term expectation, meaning the platform should maintain a non-negative income over the long term to ensure its sustainability.

The central issue at hand is not simply whether a reward function adheres to OCD-IC and BB, but rather under what conditions it meets these criteria, given the varying scenarios of task payments and opportunity costs. Consider two scenarios involving a single miner, where the same reward mechanism leads to divergent outcomes.

\begin{example}
\label{eg: 1}
    In each round, the platform garners \$100 from task payments. The miner would receive \$50 if dedicating her full computation power to alternative profits, such as Bitcoin mining.
\end{example}

\begin{example}
\label{eg: 2}
    In each round, the platform garners \$50 from task payments. The miner would receive \$80 if dedicating her full computation power to alternative profits, such as Bitcoin mining.
\end{example}

In Example \ref{eg: 1}, the platform's sufficient task payment income allows for offering miners a reward surpassing alternative sources, thus easily ensuring both OCD-IC and BB. However, Example \ref{eg: 2} presents a scenario where the platform must provide a \$30 subsidy to the miner each round to sacrifice BB and uphold OCD-IC, or else the miner might shift her computation power away for higher profits elsewhere. These contrasting examples highlight the importance of designing a reward mechanism for crowd-sourced computing platforms that minimizes reliance on assumptions about external, uncontrollable factors, particularly opportunity costs. Consequently, the primary question this paper addresses is:

\begin{question}
    \textit{``How to design a reward mechanism that minimizes assumptions on task payments and opportunity cost while still fulfilling (D)OCD-IC and BB?''}
\end{question}

\subsection{Our Results}

\noindent\textbf{Game-Theoretical Model for Crowd-Sourced Computing Platforms.}\quad
This work constructs the incentive issue as a repeated-game reward mechanism, with the dual aims of motivating miners to commit their entire computation power to the platform and ensuring the platform's budget balance. It quantifies miners' strategic decisions using a continuous value, representing the fraction of computation power they decide to invest in the platform. The influx of tasks in each round is quantified by a metric termed \textit{total difficulty}, conceptualized as a continuous random variable originating from external factors. Accordingly, the platform's task payments correlate linearly with the total difficulty, whereas the miners' opportunity cost is a function of their computation power, equating to potential earnings from alternative ventures outside the platform. Furthermore, the model treats miners' computation transcripts as a random variable with a Gamma distribution relative to their computation power, accounting for the unpredictability in task processing and hardware performance. The platform's objective is to craft a reward mechanism that considers computation transcripts, task payments, and each miner's total computation power. Section \ref{sec: model} will precisely define this model, and Section \ref{reward function section} will formally define the three critical requirements for the reward mechanism: Opportunity-Cost-Driven Incentive Compatibility (OCD-IC), Dynamic OCD-IC (DOCD-IC), and long-term Budget Balance (BB).

\vspace{0.5em}
\noindent\textbf{Analysis on Conventional Pay-Per-Share (PPS) Reward Mechanism.}\quad
After establishing the model, Section \ref{sec: PPS} delves into a theoretical examination of the conventional Pay-Per-Share (PPS) reward mechanism, a common choice among crowd-sourced platforms. Theorem \ref{thm:1} confirms the natural budget balance of the PPS reward, while Theorems \ref{thm:2} and \ref{thm:3} reveal that achieving OCD-IC necessitates assumptions regarding both task payments, i.e., demand, and opportunity costs. Typically, the PPS mechanism presupposes that demand exceeds supply and that the opportunity cost has an upper bound. These findings indicate the challenges in sustaining long-term miner contributions without accounting for opportunity costs. However, we show that the PPS mechanism fails to satisfy DOCD-IC in any situation in \ref{thm:4}.

\vspace{0.5em}
\noindent\textbf{A Novel Pay-Per-Share with Subsidy (PPSS) Reward Mechanism.}\quad
Acknowledging the limitations of the conventional Pay-Per-Share (PPS) reward mechanism, this paper suggests modifying the budget balance criteria for platforms. To ensure sustainability, achieving a long-term budget balance—where the platform may subsidize miners under certain conditions without jeopardizing its long-term expected non-negative income—is deemed sufficient. Section \ref{sec: PPSS} introduces the innovative Pay-Per-Share with Subsidy (PPSS) reward mechanism, designed to bypass the need for assumptions about opportunity cost, thereby fulfilling both OCD-IC and long-term BB as demonstrated in Theorem \ref{thm:4} \& \ref{thm:5}. This reward mechanism takes into account miners' historical computation transcripts as a reflection of their genuine computation power input and integrates the opportunity cost function for a precise and just reward allocation. While it still predicates on a demand surpassing supply, it completely dispenses with the assumption on opportunity cost, rendering the mechanism more adaptable and universally applicable to various real-world crowd-sourced computing platforms. Furthermore, in Theorem \ref{thm:7}, we investigate the conditions under which the PPSS mechanism satisfies DOCD-IC.

\subsection{Related Work}
%crowdsouring mechanism deisgn (Hong)
\textbf{Crowd-sourced Mechanism Design.}\quad
The field of crowd-sourced mechanism design is witnessing rapid growth, mirroring the burgeoning interest in leveraging collective efforts for diverse tasks. Singla et al. \cite{Singla} lead with innovative solutions for optimal pricing in crowdsourcing, blending procurement auctions with multi-armed bandit techniques to ensure budget feasibility, near-optimal utility for requesters, and incentive compatibility for workers. Anari et al. \cite{Nima2014} advance the discourse with a budget-feasible mechanism that attains an unmatched competitive ratio of 1 - 1/e, enhancing requester utility in large markets and accommodating submodular utility functions for wider applicability. Liu et al. \cite{liu2014crowdsourcing} provide empirical insights from the Taskcn platform, showing that increased rewards significantly uplift both the quantity and quality of submissions, albeit with the caveat that early high-quality submissions might suppress further quality contributions. Easley et al. \cite{Easley2015} critique traditional incentive models by integrating prospect theory into crowdsourcing markets, illustrating that custom incentives can outperform fixed-payment schemes by better matching actual worker behaviors. Shah et al. \cite{Shah2016} introduce a distinctive payment mechanism to improve data quality by encouraging workers to select tasks they're confident about, thus minimizing errors and spam. Fan et al. \cite{Fan2020} address risk distribution in micro-task crowdsourcing, promoting a mechanism that ensures a standard hourly wage to motivate the undertaking of complex tasks without fear of uncompensated effort. Yang et al. \cite{YangCC2021} explore gamification in crowdsourcing, examining the impact of intrinsic and extrinsic motivators, such as points and immediate feedback, on user engagement and behavior. Truong et al. \cite{truong2023efficient} reveal an innovative algorithm for selecting incentives in crowdsourcing endeavors, viewing the issue as an online decision-making problem to maximize utility while adhering to budget and time constraints. Luo \cite{luo2023incentivizing} presents the Effort Incentivization (EI) mechanism, capitalizing on peer effects to ensure worker actions align with project objectives, markedly reducing costs and enhancing task completion rates for projects with interconnected micro-tasks. Pakzad-Hurson \cite{pakzad2023crowdsourcing} devises a mechanism that approximates optimal allocations under imperfect information, employing a "wisdom-of-the-crowd" approach to penalize deviations and achieve near-optimal outcomes, demonstrating its versatility across various domains. These contributions underscore the dynamic evolution of mechanism design in crowd-sourcing, bridging theoretical constructs with practical application. Our study distinguishes itself by introducing a novel incentive framework specifically crafted for crowd-sourced computing to adeptly address the balance between participant opportunity costs and platform fiscal health.

\vspace{0.5em}
\noindent\textbf{Opportunity Cost Consideration.}\quad
%oppotunity cost related (Hong)
Opportunity cost plays a pivotal role in economic theory, profoundly impacting mechanism design, particularly in settings where agents face choices among various alternatives. This section synthesizes essential studies that incorporate opportunity cost into mechanism designs, thereby increasing participant engagement and aligning these systems more closely with real-world conditions.
Lu \cite{lu2009auction} explores revenue-maximizing auctions, highlighting the importance of acknowledging bidders' opportunity costs. He advocates for second-price sealed-bid auctions with tailored reserve prices and entry subsidies, underlining the need for auction variety, strategic participant number restrictions, and the advantages of asymmetric entry, establishing a foundation for effectively integrating opportunity costs.
Read et al. \cite{read2017value} examine the effects of opportunity cost framing on patience, finding that emphasizing the immediate costs associated with smaller, sooner rewards fosters increased patience. This insight aids in comprehending the behavioral implications of awareness of opportunity costs.
Gao et al. \cite{gao2017new} introduce a budget allocation model based on expected opportunity cost (EOC), facilitating a decision-making paradigm that significantly penalizes suboptimal selections. Tailored for risk-neutral decision-makers, this framework aligns well with the complex dynamics surrounding opportunity costs.
Fleckenstein et al. \cite{fleckenstein2023concept} address the estimation challenges of opportunity costs within the interconnected realms of demand management and vehicle routing, highlighting the crucial balance between demand decisions and logistical planning.
Zhang et al. \cite{zhang2023contract} evaluate buyback and revenue-sharing contracts through the lens of the opportunity cost of working capital (OCWC), revealing how these costs influence the attractiveness and selection of contract types, offering strategic insights for choosing between contracts.
Tyagi et al. \cite{tyagi2023riggs} devise auction protocols that reduce participants' opportunity costs by adapting auction dynamics and collateral requirements to accommodate bid withdrawals, proposing solutions such as early collateral release and mutual consent for auction termination to minimize these costs.
Goyal et al. \cite{Goyal_ec} refine Constant Function Market Makers (CFMMs) in the decentralized finance sector by developing trading functions that consider liquidity providers' opportunity costs, fee revenue, and arbitrage losses. This innovation leads to CFMM strategies that more accurately mirror market-maker forecasts and operational compromises.
In light of this discussion, our research presents a game-theoretical framework for crowd-sourced computing platforms, tackling the unique challenges presented by strategic miners potentially diverting their computational efforts to more lucrative activities like Bitcoin mining. We introduce Opportunity-Cost-Driven Incentive Compatibility (OCD-IC) to maintain miner engagement by compensating for possible alternative earnings. Detailed in Section~\ref{reward function section}, this method signifies our effort to weave opportunity cost considerations into incentive mechanisms for crowd-sourced platforms, ensuring they resonate with miners' economic motives and the sustainability of the platform.

\vspace{0.5em}
%Bitcoin mining pool (Ziyue)
\noindent\textbf{Mining Pool Reward Mechanism Design.}\quad
This study draws inspiration from recent research on blockchain mining pool reward mechanisms \cite{IncentiveCompatibilityBitcoin,variance,rosenfeld2011analysis,eyal2015miner,fisch2017socially,romiti2019deep}. Rosenfeld \cite{rosenfeld2011analysis} outlines various Bitcoin pooled mining reward functions, such as Pay-Per-Share and Pay-Per-Last-N-Share. Fisch et al. \cite{fisch2017socially} develop a formal model using discounted expected utility to assess the effectiveness of pooling strategies in general crowd-sourced pool mining, proposing the geometric pay pool to ensure steady-state utility among miners. Roughgarden et al. \cite{variance} showed that adding complexity to pool reward distribution by categorizing shares into multiple types increases reward variance for miners and highlighted that Pay-Per-Share rewards minimize this variance, ensuring long-term rewards are proportional to each miner's computing power. Eyal \cite{eyal2015miner} investigates sabotage attacks in open mining pools through a game-theoretic lens, demonstrating that pools' strategies of infiltrating each other to withhold proofs of work result in a non-cooperative equilibrium detrimental to all. Addressing this issue, Schrijvers et al. \cite{IncentiveCompatibilityBitcoin} suggest an incentive-compatible reward mechanism in Bitcoin mining pools that discourages miners from delaying solution submissions, ensures payments proportional to computing power, and maintains budget balance. Diverging from previous research, our methodology conceptualizes computing power as a compact set and explores a reward mechanism aiming to motivate miners to dedicate the entirety of their computing power to our platform, thereby promoting the pool's sustainability and stability.

\section{Preliminaries}
\textbf{Notation and Terminology.}\quad
In the paper, miners are indexed by $i$, and rounds are indexed by $j$. The strategy space available to miners is denoted as $\mathbf{A}=(\mathsf{A}_1, \mathsf{A}_2, \ldots, \mathsf{A}_n)$, wherein each $\mathsf{A}_i=[0,A_i]$ is a compact set indicating all the pure strategies accessible to miner. The collection of strategies adopted by all miners is denoted by $\mathbf{a}\in\mathbf{A}$. Regarding vector operations, the expression $\mathbf{X}_1 + \mathbf{X}_2$ signifies the element-wise summation of vectors $\mathbf{X}_1$ and $\mathbf{X}_2$, while $\vert \mathbf{X} \vert$ corresponds to the Manhattan norm of vector $\mathbf{X}$.

\vspace{0.5em}
\noindent\textbf{Computing Model.}\quad
We assume the computational model follows the Gamma distribution, characterized by a shape parameter that is a function of the computing power $a_i$ employed by miner $i$, adjusted by a constant multiplier $k$, and a rate parameter fixed at $1$. We selected the Gamma distribution for its flexibility and suitability in modeling positive-valued, normally distributed computational difficulty. This framework aims to capture the inherent randomness of computational tasks associated with mining, emphasizing the direct correlation between the resolution of difficulties and the applied computing power. The adoption of the Gamma distribution serves not only as a versatile and coherent framework to illustrate the inherent variability and unpredictability of these tasks but also offers intuitive insight that an increase in computing power proportionally enhances the outcomes. In this paper, $D_i$ represents the tally of difficulties processed by miner $i$. Specifically, the formula $\mathbb{E}[D_i] = k \cdot a_i$ quantifies the expected volume of difficulties that can be addressed, showcasing the linear correlation in expectation between computing power and the outcomes of mining. This shows the rule of thumb that increasing computing power directly contributes to a greater probability of effectively addressing computational challenges. For a collection of $n$ miners, this notation is expanded as follows:
\[
\mathbf{D}^\mathbf{a} \sim \text{Gamma}(k \cdot \mathbf{a}, 1) \quad \text{for} \quad \mathbf{a} = (a_1, a_2, \ldots, a_n).
\]
Furthermore, \(\sigma(\mathbf{D}^\mathbf{a})\) is introduced to denote the sigma algebra generated by the random vector \(\mathbf{D}^\mathbf{a}\).

\vspace{0.5em}
\noindent\textbf{Reward mechanism.}\quad
Within our model, the reward mechanism to each miner $i$ is represented by the function $R_i$, denoted as $\sigma(\mathbf{D}^\mathbf{a})$. Formally, we define the reward for miner $i$ as:
\[
R_i(\mathbf{D}^\mathbf{a}) : \sigma(\mathbf{D}^\mathbf{a}) \rightarrow \mathbb{R},
\]
where $\mathbf{D}^\mathbf{a}$ states the combined computational achievements of all miners, aligned with their strategic choices $\mathbf{a} = (a_1, a_2, \ldots, a_n)$, and each strategy $a_i$ is a member of the action space $\mathsf{A_i}$. The function $R_i$ assesses the transcript $\mathbf{D}^\mathbf{a}$, which serves as evidence of the difficulty completed by the miners for the determination of rewards. Given that each miner's strategy remains confidential and is not disclosed to the platform, it necessitates that the reward distribution relies on the observable transcript $\mathbf{D}^\mathbf{a}$. This approach allows for equitable reward allocation without requiring access to the private strategic choices of the miners $a_i$.
\section{The Model}

\label{sec: model}

We consider a scenario with $n\geq 2$ miners engaged in computing a pre-specified total difficulty $M_j>0$ for each round $j$, where each unit of difficulty is assigned a price $p$. Consequently, the aggregate reward available for distribution in each round equals $p\cdot M_j$. The total difficulty $M_j$ for each round is selected from a distribution $F$, with the platform having access to the value of $M_j$. The mean number of difficulty units, denoted by $\mu_F$, is given by $\mu_F=\mathbb{E}[M_j]$. The rationale for assuming a continuous distribution for $F$, representing the total difficulty in each round, is underpinned by the extensive nature of computing challenges in mining processes, typically comprising millions of discrete subtasks. This vast scale facilitates the approximation of discrete events with a continuous distribution, enhancing the model's analytical tractability without significant loss of precision. The continuum assumption of difficulty number simplifies the mathematical analysis and offers a more universal framework for examining the distribution of computing tasks. Moreover, this assumption is practical and mirrors real-world scenarios where the granularity of tasks is so fine that discrete and continuous models converge in terms of their predictive accuracy and utility.

Each miner, indexed by $i$, is equipped with a total computing capacity $\mathsf{A}_i$, with the actual computing power deployed by miner $i$ being $a_i\in\mathsf{A}_i$. In this study, the opportunity cost incurred by miner $i$ for allocating $a_i$ units of computing power is represented by $C(a_i)$. This model recognizes the incremental increase in opportunity cost that arises from allocating additional computing power to the mining effort. Consistent with standard economic theory, the opportunity cost function is characterized as continuous, convex, and strictly increasing in relation to the provided computing power. For instance, the cost function can be expressed as $C(x)=r\cdot x$, where $r$ represents a constant positive rate that is independent of the platform's operations.

In this paper, miner $i$ chooses the fixed strategy $\mathbf{a}\in\mathsf{A}_i$ and submits her achieved difficulties $D_i$ to the platform in each round \(j\). This framework assumes that miners are unable to falsify their reported difficulties. The collective results from a joint strategy $\mathbf{a} \in \mathbf{A}$ in round $j$ are denoted as $\mathbf{D}_j^\mathbf{a}$. In response to the platform's established reward mechanism $R$, each miner $i$ evaluates $R$ and chooses a strategy $a_i \in \mathsf{A}_i$ that maximizes their payoff for that round, i.e.,
\[
P_i(\mathbf{a}; R) = R_i(\mathbf{D}_j^\mathbf{a}) - C(a_i), ~\forall \mathbf{a} \in \mathbf{A},~\mathbf{D}_j^\mathbf{a} \sim \text{Gamma}(k \cdot \mathbf{a}, \mathbf{1}).
\]
Given the stochastic nature of the computing process, encapsulated as a gamma random variable, it is pertinent to assess the reward mechanism through the lens of expected payoffs for each miner. The emphasis on expected values is instrumental for miners to discern strategies that yield consistent and predictable earnings in the participation of cost function $C_i$, i.e., \(~\forall \mathbf{a}\in\mathbf{A},~ \mathbf{D}^\mathbf{a}\sim\text{Gamma}(k\cdot \mathbf{a},\mathbf{1})\)
\[ 
\mathcal{P}_i(\mathbf{a}; R)=\mathbb{E}[P_i(\mathbf{a}; R)]=\mathbb{E}[R_i(\mathbf{D}^\mathbf{a})]-C(a_i).
\]

\section{Reward Function Desiderata}
\label{reward function section}
In our model, we introduce several key properties essential for an effective reward mechanism, with one of the initial considerations being the principle of budget balance. This property is crucial to ensure the financial viability of the platform, stipulating that the distribution of rewards to miners in each round is managed in a way that maintains the platform's budget health.
\begin{dfn}[Budget Balanced]
A reward mechanism $R$ is termed $(\theta,\gamma)$-budget balanced if it satisfies the condition that, for any given round $j$ and for every strategy profile $\mathbf{a} \in \mathsf{A}$,
\[\theta\leq \sum_i \frac{R_i(\mathbf{D}_j^\mathbf{a})}{M_j\cdot p}\leq \gamma.\] 
Specifically, a reward function that is $(\theta,1)$-budget balanced directly ensures that the platform does not incur a deficit in any round of the mining process.
\end{dfn}
Achieving budget balance in every round, although ideal for financial stability, may only sometimes be realistic due to the stochastic nature of demand and the inherent randomness in computational efforts. These unpredictable elements can make maintaining a perfectly balanced budget in each round challenging. Consequently, we introduce the notion of long-term budget balance becoming necessary and meaningful. This relaxed approach accommodates the fluctuations in miner participation and computational contributions, which is crucial for platforms subject to the dynamic nature of crowded-sourced computing. It enables the platform to manage temporary imbalances with an overarching goal of achieving fiscal stability over time. 
\begin{dfn}[long-term Budget Balanced]
A reward mechanism $R$ is said to achieve $(\theta,\gamma)$-long-term budget balance if, across all possible joint strategies $\mathbf{a} \in \mathsf{A}$,
\[\theta\leq \sum_i \mathbb{E}\left(\frac{R_i(\mathbf{D}^\mathbf{a})}{M_j\cdot p}\right)\leq \gamma.\]
\end{dfn}

We then introduce Opportunity-Cost-Driven Incentive Compatibility (OCD-IC). This property is pivotal in aligning the payoff of miners with the objectives of the platform. The essence of OCD-IC lies in its design to encourage miners to allocate their maximum computing power to the platform, factoring in the presence of alternative opportunities or costs. This consideration is crucial for the platform, which aims to maximize the computational resources at its disposal while ensuring miners are adequately compensated for their contributions.

\begin{dfn}[Opportunity-Cost-Driven Incentive Compatible]
\ \ A reward function $R$ is defined as Opportunity-Cost-Driven Incentive Compatible (OCD-IC) if it ensures that for every miner $i$, the optimal allocation of computing power to maximize their payoff corresponds to their total available capacity $A_i$. Formally, this condition is met if, for all $i$,
\[\arg\max_{a\in [0,A_i]}\floor*{\mathcal{P}_i(a; R)}=A_i.\]
\end{dfn}
The term "Opportunity-Cost-Driven" underscores the model's acknowledgment of the real-world scenario where miners face decisions that involve trade-offs. Miners are more likely to dedicate their computing resources to the platform when the reward mechanism is structured to account for these opportunity costs, thus maximizing their payoff. By focusing on this property, the designed reward mechanism not only ensures that the platform attracts the necessary computational power for its operations but also establishes a fair and competitive environment where each miner's best strategy is to fully engage with the platform's demand. In order to further examine the properties of the reward mechanism in a multi-round game, we introduce a definition that focuses on its behavior in each individual round. Specifically, this definition captures how the reward mechanism influences miners' short-term decisions within a single round.
\begin{dfn}[Dynamic OCD-IC]
A reward function \( R \) is defined as Dynamically Opportunity-Cost-Driven Incentive Compatible (DOCD-IC) if, in each round \( j \), it ensures that for every miner \( i \), the optimal allocation of computing power to maximize their immediate payoff corresponds to their total available capacity \( A_i \). Formally, this condition is met if, for all \( i \) and for each round \( t \),
\[
\arg\max_{a \in [0, A_i]} P_{i,j}(a; R)  = A_i,
\]
\( P_{i,j}(a; R) \) denotes the miner's payoff in round \( j \) given the reward function \( R \).
\end{dfn}

\section{Pay-Per-Share Reward Mechanism}
\label{sec: PPS}
In this section, we conduct a theoretical exploration of the Pay-Per-Share (PPS) reward mechanism, a prevalent model within Bitcoin mining pools for the distribution of earnings. This mechanism allows miners to contribute their computational power, referred to as "hash rate,"\cite{variance} towards the resolution of intricate cryptographic challenges. A miner generates a "share" upon discovering a solution that meets the pool's difficulty criteria, albeit potentially falling short of the Bitcoin network's broader requirements. Compensation for each share is predetermined, offering a set amount of bitcoin regardless of the pool's success in block discovery,
\[R^{pps}_i(X_i)=X_i\cdot c,\]
where $X_i$ represents the count of "shares" submitted by miner $i$ and $c$ stands as a fixed parameter, established beforehand and known to all participating miners. The PPS model, extensively analyzed in prior studies\cite{variance,rosenfeld2011analysis,IncentiveCompatibilityBitcoin}, is favored for its direct correlation between rewards and individual miner hash rates, effectively minimizing reward variance and facilitating stable, predictable earnings for miners. Furthermore, it incentivizes the timely submission of solutions, enhancing the potential for expedited block generation. Nonetheless, it has been observed that the PPS scheme may not always achieve budget balance in every round within a Bitcoin mining pool, suggesting the possibility of deficits under certain circumstances. We extend our analysis to the PPS reward mechanism within the context of a crowd-sourced computing platform. For each round $j$, a fixed reward per unit of difficulty is established as $b$. Under the Pay-Per-Share framework for a given strategy $\mathbf{a} \in \mathsf{A}$, the mechanism is designed as follows:
\[R^{pps}_i(\mathbf{D}_j^\mathbf{a}) = 
\begin{cases} 
\frac{D_i}{|\mathbf{D}_j^\mathbf{a}|} \cdot b \cdot \min\{|\mathbf{D}_j^\mathbf{a}|, M_j\} & \text{if } |\mathbf{D}_j^\mathbf{a}| \neq 0, \\
0 & \text{if } |\mathbf{D}_j^\mathbf{a}| = 0.
\end{cases}\]
\begin{theorem}
\label{thm:1}
The Pay-Per-Share (PPS) reward mechanism $R^{PPS}$ maintains a $\left(0,\frac{b}{p}\right)$-budget balance for each individual round $j$, and by extension, PPS mechanism satisfying $\left(0,\frac{b}{p}\right)$-long term budget balance.
\end{theorem}
The theorem \ref{thm:1} states that the Pay-Per-Share (PPS) reward mechanism maintains a budget balance independent of the total difficulty distribution $F$. Furthermore, setting $b=p$ to achieve a $(0,1)$-budget balance is a pivotal safeguard against deficits. By ensuring the equitable distribution of all collected funds back to the miners, this approach prevents fiscal shortfalls of the platform. Additionally, the PPS mechanism's provision of immediate rewards for each share submitted significantly fosters a stable community of miners. This model of immediate compensation encourages ongoing participation, offering miners consistent and predictable rewards. The resultant stability benefits both the miners through reliable income and the platform by securing a continuous supply of computational power necessary for its operations. The proof of theorem \ref{thm:1} is shown below.

In the subsequent analysis, we explore the Opportunity-Cost-Driven Incentive Compatibility (OCD-IC) of the Pay-Per-Share (PPS) reward mechanism. Initially, we examine a simple scenario characterized by a linear cost function, $C(x) = r \cdot x$, where $r \geq b \cdot k$. Under this assumption, it is possible that $R^{PPS}$ does not satisfy the criteria for OCD-IC. This is attributed to the comparative analysis of the cost function $C(a_i)$ against the potential rewards from $R^{PPS}_i$, where the opportunity costs or the computational expenses may surpass the rewards provided by the PPS mechanism, potentially leading to a reduced incentive for miners to dedicate their computational resources to the platform.

\begin{theorem}
\label{thm:2}
Let the average number of difficulty, $\mu_F$, satisfy $\mu_F \geq k \cdot \sum_m A_m$. For a linear cost function $C(x) = r \cdot x$ with $r \geq b \cdot k$, the Pay-Per-Share (PPS) reward mechanism $R^{PPS}$ is not OCD-IC. Conversely, if $r \leq b \cdot k$, then $R^{PPS}$ is OCD-IC.
\end{theorem}
The theorem presented reveals a compelling aspect of the Pay-Per-Share (PPS) reward mechanism's interaction with miners' cost structures and incentive alignment. Specifically, the distinction drawn by the parameter relationship $r > b \cdot k$ versus $r \leq b \cdot k$ encapsulates a critical threshold in determining miners' optimal strategies regarding the allocation of their computational resources.

When the cost coefficient $r$, indicative of the opportunity cost of computing per unit, exceeds the product of the reward per unit of computing power $b\cdot k$, the theorem suggests that miners' optimal strategy shifts towards $a_i = 0$. This outcome intuitively aligns with economic rationality; if the cost of contributing computational power surpasses the potential rewards, miners are disincentivized from participating, favoring external opportunities or avoiding the higher costs associated with mining.

Conversely, when $r \leq b \cdot k$, the conditions foster an environment where the full allocation of computational resources, $a_i = A_i$, becomes the optimal strategy for miners. This scenario underscores a situation where the rewards from the PPS mechanism sufficiently compensate for or exceed the costs incurred by miners, thereby motivating them to contribute their maximum computing capacity.

The assumption $\mu_F\geq k\cdot \sum_m A_m$ implies the condition that demand for computing is greater than supply by the crowded-sourced platform. This can simplify our analysis of the PPS reward mechanism. Without this assumption, in Theorem 5.3, the bound for r will be tighter and consequently too complicated to analyze effectively.

The premise that the need for computational resources surpasses the available supply finds strong justification when examining the broad and varied sectors dependent on extensive computational inputs. Domains like artificial intelligence and machine learning lead the charge, requiring enormous processing capacity for the development and training of advanced deep learning frameworks to sift through and assimilate information from large data volumes\cite{Strubell_Ganesh_McCallum_2020}. In a similar vein, the areas of climate modeling and weather prediction engage in complex simulations of environmental dynamics, dependent on significant computational strength for precise pattern forecasts. In cryptocurrency mining, the necessity for specialized machinery to decode cryptographic challenges highlights the severe requirement for computational resources. Furthermore, in the life sciences, especially within genomics and bioinformatics, the processing of extensive genetic data sets challenges the limits of current computational infrastructures. Financial modeling also intensifies the need for computational power through its dependence on algorithms for high-frequency trading and risk evaluation, demanding swift and comprehensive data analysis. Likewise, the sectors of computer graphics, big data analysis, and pharmaceutical research uniformly necessitate substantial computational endeavors to produce intricate animations, derive insights from voluminous data collections, and model molecular dynamics, respectively. Each of these areas illustrates not only the extensive need for computational power but also the challenges in meeting these demands; thereby, it is reasonable to assume the demand for computing is greater than the supply.

The next theorem explores the more general assumptions so that the PPS mechanism can be OCD-IC but it is not DOCD-IC in any case.
\begin{theorem}
\label{thm:3}
Let each miner $i$ have a cost function $C_i(x)$ that is continuous, convex, and strictly increasing, satisfying $C_i(0) = 0$ over the action space $[0, A_i]$, and let $\mu_F \geq k \cdot \sum_m A_m$. The Pay-Per-Share (PPS) reward mechanism $R^{PPS}$ is Opportunity-Cost-Driven Incentive Compatible (OCD-IC) if and only if for all miners $i$, the marginal cost $\frac{\partial C(a)}{\partial a}$ satisfies $\frac{\partial C(a)}{\partial a} \leq b \cdot k$ for all $a \in [0, A_i]$.
\end{theorem}

\begin{theorem}
\label{thm:4}
    The Pay-Per-Share mechanism is not DOCD-IC.
\end{theorem}

Inspired by the reward functions used in Bitcoin mining pools, this section explored the budget balance of the Pay-Per-Share (PPS) reward mechanism and its Optimal Cost-Driven Incentive Compatibility (OCD-IC) in the context of crowd-sourced computing platform. Our findings illuminate the nuanced behavior of the PPS mechanism, showcasing its ability to maintain budget balance as detailed in Theorem \ref{thm:1}. However, it becomes evident that the mechanism does not universally guarantee OCD-IC. Examining various scenarios shaped by the miners' cost function configurations shows that OCD-IC is achievable under certain precise conditions outlined in Theorems \ref{thm:2} and \ref{thm:3}. These conditions, albeit rigorous, may not consistently align with the heterogeneous and dynamic nature of practical, crowded-soured computing platforms. Also, we show that the mechanism is not DOCD-IC. In light of this realization, next section introduces an alternative reward function designed to hold the OCD-IC property intrinsically.

\section{Pay-Per-Share with Subsidy reward mechanism}
\label{sec: PPSS}
In this section, we unveil the Pay-Per-Share with Subsidy (PPSS) reward mechanism, a new approach designed to inherently possess the Opportunity-Cost-Driven Incentive Compatibility (OCD-IC) property under the premise of convex, increasing, and continuous cost functions. This model necessitates an augmented capability on the part of the platform operator, specifically, the ability to discern the cost function $C_i$ pertinent to each miner and to ascertain the total computing power, $A_i$, allocated by each participant $i$. The intuition of the PPSS mechanism is discussed through three key concepts:

\begin{itemize}
    % \vspace{-0.5em}
    \item \textbf{Guaranteed Minimum Reward:} Mirroring the Pay-Per-Share reward mechanism in the cryptocurrency mining pool, the PPSS mechanism is designed with the basic compensation for every miner's computing power. This base reward is critical in ensuring that participants are consistently compensated for their contributions.
    
    \item \textbf{Who Can Get the Subsidy:} With the platform operator's enhanced information into each miner's total computing power, it becomes feasible to customize subsidy thresholds tailored to the computing capacity of each miner, thereby preserving equity among participants. For example, a miner possessing a higher computational power, denoted as $A_1$, would be assigned a higher subsidy threshold compared to a miner with a lower capacity, $A_2$. This differential approach to subsidy thresholds is instrumental in maintaining a fair incentive structure.
    
    \item \textbf{How is the Subsidy Calculated:} The subsidy component is crucial for incentivizing miners to allocate most of their computing resources to the platform. However, since cost functions $C_i$ are different, the platform must design the subsidy carefully based on its own cost function. While there is a potential risk of miners misrepresenting their cost functions to gain higher rewards, this paper assumes miners will report their \(C_i\) accurately. The exploration of strategic misrepresentation forms the basis for future research. Moreover, the disparity in computing power among miners necessitates that the subsidy computation accounts for each miner's capacity ($A_i$), ensuring that the rewards are proportionately adjusted based on contributions relative to expected benchmarks.
\end{itemize}

\vspace{-0.2em}
For strategy $\mathbf{a} \in \mathsf{A}$, the Pay-Per-Share with Subsidy reward mechanism in round $j$ is defined as

\[
\scalebox{0.7}{
$R^{PPSS}_i(\mathbf{D}_j^\mathbf{a}) = 
\begin{cases} 
\frac{D_i^j}{|\mathbf{D}_j^\mathbf{a}|}\cdot\left(b+B_i(N,\lambda)\cdot\frac{\frac{\tilde{c}_i}{k}-b}{K_i(D_i^j, \lambda)}\right)\cdot \min\{|\mathbf{D}_j^\mathbf{a}|,M_j\} & \text{if } |\mathbf{D}_j^\mathbf{a}| \neq 0, \\
0 & \text{if } |\mathbf{D}_j^\mathbf{a}| = 0.
\end{cases}$
}
\]
\[
\scalebox{0.8}{
$B_i(N, \lambda) = 
\begin{cases} 
1 & \text{if} \sum_{x=j-N}^{j} \lambda\cdot D_i^x \geq A_i \cdot N \cdot k, \\
0 & \text{otherwise}.
\end{cases}$
}
\]
\[
\scalebox{0.8}{
$\tilde{c}_i=C_i'(A_i)\text{, and } K_i(D_i^j, \lambda)=1-\frac{\lambda\cdot A_i\cdot k}{D_i^j}\cdot e^{1-\frac{\lambda\cdot A_i\cdot k}{D_i^j}}.$
}
\]

\begin{figure}[htbp]
\centering
\includegraphics[width=0.5\textwidth, trim=1cm 0.4cm 1cm 0.5cm, clip]{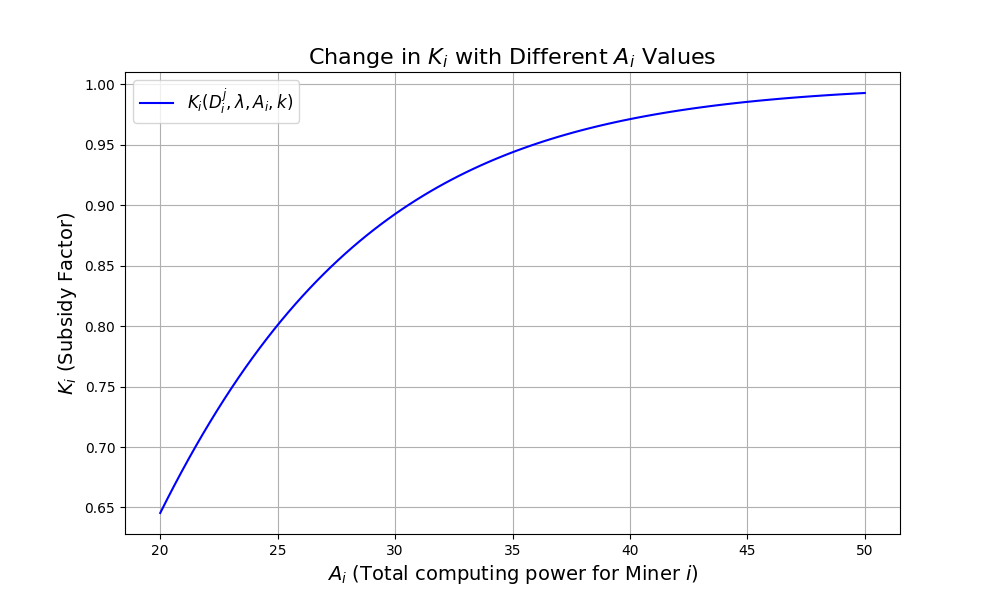}
\caption{Change in subsidy factor ($K_i$) as a function of total computing power ($A_i$) with parameters $\lambda = 0.8$, $k = 2$ when the completed difficulty $D = 10$.}
\label{fig:g1}
\end{figure}

The notation $D_i^j$ denotes the difficulty completed by miner $i$ in round j. $b$ represents a base reward for each difficulty unit, which ensures a guaranteed minimum compensation for participation, aligning with the traditional PPS model to provide a stable income for miners.
The term $B_i(N,\lambda)$ acts as an indicator function parameterized by $N$ and $\lambda$, which are both determined by the platform operator. This binary indicator function evaluates to $1$ if the miner's total contribution over $N$ rounds is at least $\lambda\cdot A_i \cdot N \cdot k$, representing the expected number of difficulty that the miner could complete in the latest $N$ rounds by allocating the most computing power. The primary purpose of incorporating $N$ is to promote sustained contribution from miners; those who consistently allocate a significant portion of their computing power to the platform are more likely to receive a subsidy.
The subsidy factor for miner $i$ is given by $\left(\frac{\tilde{c}_i}{k}-b\right)/K_i(D_i^j, \lambda)$, incorporating miner $i$'s cost function $C_i$, total computing power $A_i$, and the platform-defined parameter $\lambda$. Note that $K_i(D_i^j, \lambda)$, which is the strictly decreasing function bounded by $(0,1]$, is influenced by $A_i$. The larger the \(A_i\), the slower the rate of decrease, indicating that for a constant \(\lambda\), if two miners solve an equal number of difficulties, the one with greater computing power receives a smaller subsidy factor. This PPSS reward mechanism establishes a clear goal for miners, motivating them to surpass a difficulty threshold that aligns with their capabilities.
Following the detailed description of the PPSS reward mechanism, we design a simple example to demonstrate the behavior of subsidy factors $K_i(D_i^j, \lambda)$. The example aims to show the PPSS reward mechanism accounts for the variance in miners' total computing capacities, denoted by \(A_i\). By integrating the capacity \(A_i\) into the subsidy factor, the reward mechanism dynamically aligns payouts with the proportional effort and computing power contributed by each miner.
\begin{example}
\label{eg: 3}
Our experiment is to analyze the subsidy factor, $K_i(D, \lambda)$, and its relationship with the computing power, $A_i$, allocated by miners. We vary $A_i$ within a specified range to observe its impact on the subsidy factor, thus evaluating the PPSS reward mechanism's ability to incentivize miners towards optimal computing power allocation. $k = 2$, $\lambda = 0.8$, $D = 10$. Varying $A_i$ from $20$ to $50$, we plot the relationship between $A_i$ and $K_i$ to analyze how different levels of computing power influence the subsidy factor in Fig. \ref{fig:g1}.
\end{example}
\begin{theorem}
\label{thm:5}
Assume that for each miner \(i\), the cost function \(C_i(x)\) is continuous, convex and strictly increasing, with \(C_n(0) = 0\) over miner $i$'s action space \([0,A_i]\), and assume $\mu_F\geq k\cdot \sum_m A_m$. PPSS reward mechanism is OCD-IC.
\end{theorem}
In our analysis, Theorems~\ref{thm:3} and~\ref{thm:4} both consider conditions under which a pay-per-share (PPS) and PPSS reward mechanism, respectively, are Opportunity-Cost-Driven Incentive Compatible (OCD-IC). As demonstrated in Theorem~\ref{thm:4}, the PPSS reward mechanism emerges from its inherent compliance with the OCD-IC under the general assumptions of cost function $C_i$ and distribution $F$. It contrasts with the requirements for the PPS mechanism outlined in Theorem~\ref{thm:3}, where an additional constraint on the cost function's marginal change with respect to action \(a\) is necessitated to ensure OCD-IC. The direct applicability of the PPSS mechanism under such generalized conditions underscores its theoretical robustness and enhances its practical attractiveness by simplifying the prerequisites for achieving incentive compatibility. By obviating the need for explicit constraints on the marginal cost of action, the PPSS mechanism facilitates a more straightforward implementation pathway to ensure that miners are incentivized to allocate their maximum computing power in the crowded-sourced computing platform.
\begin{theorem}
\label{thm:6}
The PPSS reward mechanism is $\left(0,\frac{\sum_i \tilde{c}_i\cdot A_i}{M_j\cdot p}\right)$-long term Budget Balance.
\end{theorem}
\begin{theorem}
\label{thm:7}
    Assume that for each miner \(i\), the cost function \(C_i(x)\) is continuous, convex and strictly increasing, with \(C_n(0) = 0\) over miner $i$'s action space \([0,A_i]\), and assume $\mu_F\geq k\cdot \sum_m A_m$. PPSS mechanism is DOCD-IC.
\end{theorem}
\section{Conclusion}
In this paper, we introduce a novel game-theoretic model for crowd-sourced computing platforms, highlighting the development of an OCD-IC criterion and a Pay-Per-Share with Subsidy (PPSS) reward mechanism to address the strategic dynamics of miners. Our analysis shows that traditional Pay-Per-Share (PPS) mechanisms fall short of ensuring long-term participation without restrictive assumptions, leading to the proposal of the PPSS mechanism, which incorporates subsidies and eliminates the need for assumptions on opportunity costs. This novel approach not only ensures incentive compatibility and long-term budget balance but also broadens the applicability to real-world platforms, setting a foundation for future research on its scalability and implementation. Our work greatly contributes to designing more effective and economically sound reward systems for crowd-sourced computing platforms.

\bibliographystyle{IEEEtran}
\bibliography{zkmining}

\appendix
\section{Proofs of Theorems}
\begin{manualtheorem}{\ref{thm:1}}
The Pay-Per-Share (PPS) reward mechanism $R^{PPS}$ maintains a $\left(0,\frac{b}{p}\right)$-budget balance for each individual round $j$, and by extension, PPS mechanism satisfying $\left(0,\frac{b}{p}\right)$-long term budget balance.
\end{manualtheorem}

\begin{proof}
To validate the $\left(0,\frac{b}{p}\right)$-budget balance of the Pay-Per-Share (PPS) reward mechanism within each round and, by extension, across the long-term, it suffices to demonstrate that $R^{PPS}$ adheres to the $\left(0,\frac{b}{p}\right)$-budget balance criteria for every round $j$ and for any strategy profile $\mathbf{a} \in \mathsf{A}$, as follows:
\[0\leq \sum_i \frac{R_i^{PPS}(\mathbf{D}_j^\mathbf{a})}{M_j\cdot p}\leq \frac{b}{p}.\]
In scenarios where each miner chooses not to contribute computational power, i.e., $a_i = 0$ for all $i$, this results in $|\mathbf{D}_j^\mathbf{a}| = 0$. Consequently, $R_i^{PPS}(\mathbf{D}_j^\mathbf{a})$ equates to $0$ for every miner $i$, affirming the lower bound of $0$. On the contrary, in cases where $|\mathbf{D}_j^\mathbf{a}| \neq 0$, the aggregate rewards dispensed can be articulated as:
\[\sum_i R_i^{PPS}(\mathbf{D}_j^\mathbf{a}) = \sum_i \frac{D_i}{|\mathbf{D}_j^\mathbf{a}|} \cdot b \cdot \min\{|\mathbf{D}_j^\mathbf{a}|, M_j\} \leq b \cdot M_j.\]
From this, it follows that:
\[\sum_i \frac{R_i^{PPS}(\mathbf{D}_j^\mathbf{a})}{M_j \cdot p} \leq \frac{b \cdot M_j}{p \cdot M_j} = \frac{b}{p},\]
thereby substantiating the upper bound of \(\frac{b}{p}\) for the PPS mechanism, and completing the proof.
\end{proof}

\begin{manualtheorem}{\ref{thm:2}}
Let the average number of difficulty, $\mu_F$, satisfy $\mu_F \geq k \cdot \sum_m A_m$. For a linear cost function $C(x) = r \cdot x$ with $r \geq b \cdot k$, the Pay-Per-Share (PPS) reward mechanism $R^{PPS}$ is not OCD-IC. Conversely, if $r \leq b \cdot k$, then $R^{PPS}$ is OCD-IC.
\end{manualtheorem}

\begin{proof}
To prove Theorem \ref{thm:2}, we first determine the expected reward $R_i^{PPS}$ for any miner $i$ given their contribution $a_i$ and for any round with contributions $\mathbf{D}_j^\mathbf{a}$. The expected reward can be represented as follows:
{\small
\[\mathbb{E}[R_i^{PPS}(\mathbf{D}_j^\mathbf{a})]=\frac{k\cdot a_i}{k\cdot\sum_{m} a_m}\cdot b\cdot \min\left\{k\cdot \sum_{m} a_m,\mu_F\right\} = k\cdot a_i\cdot b.
\]
}

Given the cost function $C(a_i) = r \cdot a_i$ and under the condition that $r \geq b \cdot k$, the analysis reveals that miners would not achieve a positive payoff by allocating any computing power. Hence $a_i = 0$ becomes the maximizing strategy for their payoff, establishing the lower bound scenario.

Conversely, if the condition is such that $r \leq b \cdot k$, implying the cost per unit of computing power is less than or equal to the reward per unit, the incentive for miners shifts. In this case, the optimal strategy for each miner is to commit their maximum available computing power, i.e., $a_i = A_i$, to the platform. This strategic allocation ensures that the expected payoff from $R^{PPS}$ outweighs or equals the incurred cost, thereby proving \(R^{PPS}\) is OCD-IC.
\end{proof}

\begin{manualtheorem}{\ref{thm:3}}
Let each miner $i$ have a cost function $C_i(x)$ that is continuous, convex, and strictly increasing, satisfying $C_i(0) = 0$ over the action space $[0, A_i]$, and let $\mu_F \geq k \cdot \sum_m A_m$. The Pay-Per-Share (PPS) reward mechanism $R^{PPS}$ is Opportunity-Cost-Driven Incentive Compatible (OCD-IC) if and only if for all miners $i$, the marginal cost $\frac{\partial C(a)}{\partial a}$ satisfies $\frac{\partial C(a)}{\partial a} \leq b \cdot k$ for all $a \in [0, A_i]$.
\end{manualtheorem}
\begin{proof}
Since the cost function is continuous, convex, and strictly increasing, $C_i(a)$ and $\frac{\partial C_i(a)}{\partial a}$ are both increasing as $a$ increases on the domain. In this context, miners are incentivized to allocate their computing power at the point where the marginal cost of contributing an additional unit of computing power matches the marginal benefit, denoted by $\frac{\partial C_i(a)}{\partial a}=b\cdot k$. This is the optimal point for miners, beyond which contributing more would result in a marginal cost exceeding the marginal benefit, reducing the expected profit $\mathcal{P}_i$. For $R^{PPS}$ to be OCD-IC, the reward for contributing an additional unit of computing power $b\cdot k$ must be greater than the marginal cost, $\frac{\partial C_i(a)}{\partial a}$ across the entire action space $[0,A_i]$ for any miner $i$. If the condition $b\cdot k\geq \frac{\partial C_i(a)}{\partial a}$ holds true for all miner $i$,
\begin{align*}
\mathcal{P}_i(a_i; R^{PPS}) &= \frac{a_i}{\sum_{m} a_m}\cdot b\cdot \min\left\{k\cdot \sum_{m} a_m,\mu_F\right\}-C(a_i) \\
&= b\cdot k\cdot a_i-C(a_i).
\end{align*}
To maximize $\mathcal{P}_i(a_i; R^{PPS})$, we take the derivative with respect to $a_i$,
\[   \frac{\partial\mathcal{P}_i(a_i; R^{PPS})}{\partial a_i}=b\cdot k - \frac{\partial C(a_i)}{\partial a_i}. \]

Therefore, the result from (1) implies that the optimal action for any miner $i$ is to set $a_i=A_i$ for maximizing $\mathcal{P}_i$, i.e. $\argmax_{a\in[0,A_i]}\mathcal{P}_i(a_i; R^{PPS})=A_i$.
\end{proof}

\begin{manualtheorem}{\ref{thm:4}}
    The Pay-Per-Share mechanism is not DOCD-IC.
\end{manualtheorem}

\begin{proof}
Consider a setting where miners have access to the total difficulty $M_j$ and a fixed reward per unit difficulty $b$ before setting strategy for round $j+1$. Under the Pay-Per-Share mechanism, if a miner earns a reward of $b\cdot D_i$ in round $j+1$, she can deduce that the total difficulty contributed by all miners in round $j$, $|\mathbf{D}_j^{\mathbf{a}}|$, was less than or equal to $M_j$. Consequently, the miner gains no extra strategic information from her reward that would enable her to manipulate her strategy for $M_{j+1}$ to increase her payoff. \\ 
However, suppose in round $j$, the miners discover that their reward is scaled down to $\delta \cdot b \cdot D_i$ where $0 < \delta < 1$. This indicates that the total difficulty $M_j$ exceeded $|\mathbf{D}_j^{\mathbf{a}}|$ in round $j$. With this knowledge, the miners can infer the scaling factor $\delta$ and may adjust their strategies for the next round based on $\delta$. This ability to modify strategies based on the observed rewards and deduced scaling factor shows that the Pay-Per-Share mechanism fails to prevent strategic manipulation under a multi-round setting, proving it is not OCD.
\end{proof}

\begin{manualtheorem}{\ref{thm:5}}
Assume that for each miner \(i\), the cost function \(C_i(x)\) is continuous, convex and strictly increasing, with \(C_n(0) = 0\) over miner $i$'s action space \([0,A_i]\), and assume $\mu_F\geq k\cdot \sum_m A_m$. PPSS reward mechanism is OCD-IC.
\end{manualtheorem}

\begin{proof}
To prove the PPSS mechanism holds OCD-IC property, first, we need to define $g_{\lambda, \tilde{c}_i}(D_i^j)=\frac{\left(\frac{\tilde{c}_i}{k}-b\right)\cdot D_i^j}{K_i(D_i^j, \lambda)}$. We can verify that \(g_{\lambda, \tilde{c}_i}(D_i^j)\) is the convex function. Given the reward function for each miner, we can write out the payoff function for miner $i$,
\[\myscale[0.8]{P_i(a;R^{PPSS}) = \left(\frac{b}{|\mathbf{D}_j^\mathbf{a}|}+B_i(N, \lambda)\cdot g_{\lambda, \tilde{c}_i}(D_i^j)\right) \cdot \min\{|\mathbf{D}_j^\mathbf{a}|,M_j\} - C(a)}\]
When \(|\mathbf{D}_j^\mathbf{a}| \neq 0\); Otherwise, \(P_i(a;R^{PPSS})=0\). $B_i(N, \lambda)$ is a binary indicator that turns on if the miner's contribution over $N$ rounds meets a predefined threshold, and $K_i(D_i^j, \lambda)$ adjusts the reward based on the contribution's efficiency. We use $\mathrm{D}_i^N=\sum_{x=j-N}^{j} D_i^x$ as the gamma random variable representing the total number of difficulty completed by $i$ in the latest $N$ rounds. By Chernoff bound, we can establish an upper bound for the probability that the total difficulty $\mathrm{D}_i^N\leq \lambda\cdot A_i \cdot N \cdot k$. Further, the expectation of the indicator function is the probability of $B_i(N, \lambda)=1$. Then, we utilize the complement rule to derive a lower bound for the probability that $\mathrm{D}_i^N$ exceeds the subsidy threshold (2).

\[
\mathbb{P}\left(\mathrm{D}_i^N \leq 
\lambda\cdot A_i \cdot N \cdot k\right)\leq \left(\frac{e}{\frac{k\cdot a_i\cdot N}{\lambda\cdot A_i\cdot N\cdot k}}\right)^{\frac{k\cdot a_i\cdot N}{\lambda\cdot A_i}}\cdot e^{-k\cdot a_i\cdot N}.
\]
{
\small
\begin{equation}
\begin{aligned}
\mathbb{E}[B_i(N, \lambda)]=\mathbb{P}\left(\mathrm{D}_i^N \geq 
\lambda\cdot A_i \cdot N\cdot k\right)\geq 1-\left(\frac{\lambda\cdot A_i}{a_i}\right)\cdot e^{1-\frac{\lambda\cdot A_i}{a_i}}.
\end{aligned}
\label{eq:1}
\end{equation}
}

Then we can analyze the lower bound for $g_{\lambda, \tilde{c}_i}(D_i^j)$ by applying Jensen's inequality,

\begin{equation}
\begin{aligned}
\mathbb{E}[g_{\lambda, \tilde{c}_i}(D_i^j)]\geq g_{\lambda, \tilde{c}_i}(\mathbb{E}[D_i^j])=\frac{(\frac{\tilde{c}_i}{k}-b)\cdot \mathbb{E}[D_i^j]}{1-\frac{\lambda\cdot A_i\cdot k}{\mathbb{E}[D_i^j]}\cdot e^{1-\frac{\lambda\cdot A_i\cdot k}{\mathbb{E}[D_i^j]}}}.
\end{aligned}
\label{eq:2}
\end{equation}

Combining the result from (\ref{eg: 1}) and (\ref{eg: 2}), the expected payoff for miner $i$ uitilizing $a_i\in[0,A_i]$ computing power under PPSS reward mechanism given $\mathcal{H}_i^j$ can be lower-bounded by \(\floor*{\mathcal{P}_i(a_i;R^{PPSS})}\), which equals to
{
\small
\begin{equation}
\begin{aligned}
a_i\cdot k&\left(b+\mathbb{P}\left(\mathrm{D}_i^N \geq 
\lambda A_i  N  k\right) \frac{\frac{\tilde{c}_i}{k}-b}{1-\frac{\lambda\cdot A_i\cdot k}{a_i\cdot k}\cdot e^{1-\frac{\lambda\cdot A_i\cdot k}{a_i\cdot k}}}\right)-C(a_i) \\
&= a_i\cdot k\cdot \left(b+\frac{\tilde{c}_i}{k}-b\right)-C(a_i) \\
&= a_i\cdot \tilde{c}_i - C(a_i).
\end{aligned}
\label{eq:3}
\end{equation}
}

Finally by taking the partial derivative of equation (\ref{eq:3}), we can conclude that:
\[
\argmax_{a\in[0,A_i]} \floor*{\mathcal{P}_i(a_i;R^{PPSS})} =  A_i.
\]
\end{proof}

\begin{manualtheorem}{\ref{thm:6}}
The PPSS reward mechanism is $\left(0,\frac{\sum_i \tilde{c}_i\cdot A_i}{M_j\cdot p}\right)$-long term Budget Balance.
\end{manualtheorem}
\begin{proof}
By independence,
{
\small
\[
0\leq\mathbb{E}\left(\sum_i R^{PPSS}_i(\mathcal{H}_j)\right)=\sum_i \mathbb{E}\left(R^{PPSS}_i(\mathcal{H}_j)\right)\leq\sum_i \tilde{c}_i\cdot A_i.
\]
}
Therefore, PPSS reward mechanism can achieve $\left(0,\frac{\sum_i \tilde{c}_i\cdot A_i}{M_j\cdot p}\right)$-long-term budget balance.
\end{proof}
\begin{manualtheorem}{\ref{thm:7}}
    Assume that for each miner \(i\), the cost function \(C_i(x)\) is continuous, convex and strictly increasing, with \(C_n(0) = 0\) over miner $i$'s action space \([0,A_i]\), and assume $\mu_F\geq k\cdot \sum_m A_m$. PPSS mechanism is DOCD-IC.
\end{manualtheorem}

\begin{proof}
   In round $j$, even miners with incomplete information will not change their strategy because the subsidy part depends on the amount of computing power they allocated previously. An increase or decrease in demand \((M_{j+1})\) will not affect their strategy.
\end{proof}

\end{document}